\documentstyle[10pt]{article}
\advance \topmargin by -\headheight
\advance \topmargin by -\headsep     
\evensidemargin \oddsidemargin
\def\br{\begin{eqnarray}}
\def\er{\end{eqnarray}}
\def\be{\begin{equation}}
\def\ee{\end{equation}}
\def\nonu{\nonumber}

\def\({\left(}
\def\){\right)}

\relax

\def\a{\alpha}

\def\d{\delta}

\def\eps{\epsilon}

\def\o{\over}

\def\pa{\partial}
\def\pr{\prime}

\def\s{\sigma}

\def\lie{{\cal G}}

\def\rlx{\relax\leavevmode}
\def\inbar{\vrule height1.5ex width.4pt depth0pt}
\def\IZ{\rlx\hbox{\sf Z\kern-.4em Z}}
\def\IR{\rlx\hbox{\rm I\kern-.18em R}}
\def\IC{\rlx\hbox{\,$\inbar\kern-.3em{\rm C}$}}
\def\one{\hbox{{1}\kern-.25em\hbox{l}}}

\begin{document}

\begin{flushright}
IFT--P.064/98\\ 
\end{flushright}

\vskip1cm
\begin{center}
{\Large \bf  Singular Non-Abelian Toda Theories}
\\
\end{center}
\vspace{1.0cm}
{\bf J. F. Gomes, F. E. Mendon\c{c}a da Silveira, A. H. Zimerman}
{\it Instituto de F\'{\i}sica Te\'orica - UNESP,
 Rua Pamplona, 145 , 01405-900, Sao Paulo - SP, Brazil}\\
 jfg@axp.ift.unesp.br, eugenio@axp.ift.unesp.br, zimerman@axp.ift.unesp.br

\vspace{0.5cm}

\noindent
{\bf G. M. Sotkov}\footnote{On leave of absence from the Institute for Nuclear
 Research and Nuclear Energy, Bulgarian Academy of Sciences, 1784, Sofia}
{\it Departamento de Campos e Part\'{\i}culas - CBPF
 Rua Dr. Xavier Sigaud, 150 -Urca
 22290-180, Rio de Janeiro - RJ, Brazil}\\
  sotkov@cbpfsu1.cat.cbpf.br

\vspace{1.0cm} 
     
$\frac{}{}$
\noindent {\bf ABSTRACT} The algebraic conditions that specific gauged
 $G/H$-WZW model
 have to satisfy in order to give rise to Non-Abelian Toda models with {\it
 non singular} metric with or without {\it torsion} are found. 
  The classical algebras of symmetries corresponding to grade 
  one rank 2 and 3 singular NA-Toda models are derived.
\vskip 1cm

The proliferation of different two dimensional 
 $G_{0}$-Toda theories (\cite{Gervais} - \cite{ora} )  addresses the question
 about their algebraic classification
in terms of a $G_{0}\subset G$ embedding and a $G$-invariant $WZNW$-model. The
simplest class of models is the Abelian Toda, which is known to be completely
integrable and conformal invariant. These models correspond to a Abelian
 subgroup
$G_{0}\subset G$ and their symmetries generate the $W_{n}$-algebra 
($n$=rank $G$).

The non-Abelian Toda models, in turn are connected to non-abelian embeddings 
$G_{0}\subset G $ and  describe a string  propagating on a specific curved 
background, containing also tachyons, dilatons $\Phi (X)$ 
and, possibly axions. Its general action is of
the form ($i,j = 1, \cdots ,D$; $\mu ,\nu = 0,1$):
\begin{eqnarray}
S=\int d^2z  \{ \( G_{ij}(X) \eta_{\mu \nu }
+\epsilon _{\mu \nu}B_{ij} (X) \) 
\partial_{\mu }X^{i}\partial_{\nu }X^{j}
- \a^{\pr} R^{(2)} \Phi(X) + Tach. \; potential \}.
\nonu 
\end{eqnarray} 

The properties  of the background metric $G_{ij }$ and of the anti-symmetric
term $B_{ij}$(torsion) depend upon the embedding $G_{0}\subset G$, where $G_{0}$ is now
non-Abelian and classify the models according to {\it singular} or 
{\it non-singular}
metrics and the presence of {\it axionic }or {\it torsionless} terms.
 The symmetries of the singular metric NA-Toda models are  described by a non-local algebra, which
corresponds to the semi-classical limit of a mixed parafermionic and
$W$-algebra structure, denoted by
$V$-algebra (see \cite{Bilal}, \cite{Gervais}, \cite{GSZ1} and \cite{GSZ2}).

Consider the $G$-invariant $WZNW$-model, describing a 2-D conformal invariant
field theory on a group manifold, with fields parametrizing the group element
\begin{eqnarray}
g(z,\bar{z})=\exp (\alpha^{a}(z,\bar{z})T^{a}),
\label{2}
\end{eqnarray}
where $[T^{a},T^{b}]=f^{abc}T^{c}$ is the Lie bracket satisfied by the
generators of the Lie algebra of $G$.

The $WZNW$-action is given by
\begin{eqnarray}
S_{WZNW}=\frac{k}{4\pi }\int d^2zTr(g^{-1}\partial gg^{-1}\bar{\partial }g)
+\frac{k}{12\pi }\int_{B}\epsilon_{ijk}
Tr(g^{-1}\partial_{i}gg^{-1}\partial_{j}gg^{-1}\partial_{k}g),
\label{3}
\end{eqnarray}
where the topological term denotes a surface integral  over a ball $B$
identified as  space-time.

The equations of motion are
\begin{eqnarray}
\bar{\partial}J=\partial \bar{J}=0,
\label{4}
\end{eqnarray}
where the currents $J=g^{-1}\partial g$ and $\bar{J}=-\bar{\partial}gg^{-1}$
satisfy the Kac-Moody algebra
\begin{eqnarray}
\{ J^{a}(x),J^{b}(y)\} =f^{abc}J^{c}(x)\delta (x-y)
+k\delta^{ab}\partial_{x}(x-y).
\nonu
\end{eqnarray}

We now seek for a field theory in the submanifold $G_{0}\subset G$ preserving
conformal invariance. In order to eliminate unwanted degrees of freedom,
corresponding to the tangent space of $G/G_{0}$, we may either implement constraints upon
specific components of $J$ and
$\bar{J}$ \cite{Balog} or, equivalently,  propose a gauge invariant action and eliminate
degrees of freedom by choice of gauge.  In this note we shall follow the second
approach.

In order to construct a gauge invariant action, we first discuss a systematic
procedure in defining the $G_{0}$ subgroup, in terms of a grading operator.

Let $Q$ be a grading operator, decomposing the Lie algebra $\lie$ into graded
subspaces, i. e.,
\begin{eqnarray}
\lie=\oplus_{i}\lie_{i},\quad \quad 
[Q,\lie_{i}]=i\lie_{i},\quad \quad 
[\lie_{i},\lie_{j}]\in \lie_{i+j}.
\label{8}
\end{eqnarray}
It is clear that $\lie_{0}$ is the subalgebra of $\lie $.

Consider now the Gauss decomposition of the group element $g\in G$, i. e.,
\begin{eqnarray}
g=N_{-}g_{0}M_{+},
\label{9}
\end{eqnarray}
where $N_{-}=\exp \lie_{<0}$, $g_{0}=\exp \lie_{0}$ and
$M_{+}=\exp \lie_{>0}$. The Toda models correspond to the co-set
$N_-\backslash G/M_+$,
 and we seek for an action invariant under
\begin{eqnarray}
g\longrightarrow g^{\prime}=\alpha_{-}g\alpha_{+},
\label{10}
\end{eqnarray}
so that $N_{-}$ and $M_{+}$ may be eliminated by choice of gauge.

Introduce gauge fields $A=A_{-} \in \lie_{<}$ and
 $\bar{A}=\bar{A}_{+}\in \lie_{>}$, 
transforming as
\begin{eqnarray}
A\longrightarrow A^{\prime}=\alpha_{-}A\alpha_{-}^{-1}
+\alpha_{-}\partial \alpha_{-}^{-1},
\quad \quad 
\bar{A}\longrightarrow \bar{A}^{\prime}=\alpha_{+}^{-1}\bar{A}\alpha_{+}
+\bar{\partial}\alpha_{+}^{-1}\alpha_{+}.
\label{12}
\end{eqnarray}

An invariant action under transformations (\ref{10}) and (\ref{12})
is 
\begin{eqnarray}
S_{G/H}(g,A,\bar{A})&=&S_{WZNW}(g)
\nonumber
\\
&-&\frac{k}{2\pi}\int dz^2 Tr\( A(\bar{\partial}gg^{-1}-\epsilon_{+})
+\bar{A}(g^{-1}\partial g-\epsilon_{-})+Ag\bar{A}g^{-1}\) ,
\nonu
\end{eqnarray}
where $\epsilon_{\pm }$ are constant generators of grade $\pm 1$, respectively.
Since the action $S_{G/H}$ is invariant, we may choose $\alpha_{-}=N_{-}^{-1}$
and $\alpha_{+}=M_{+}^{-1}$, so that
\begin{eqnarray}
S_{G/H}(g,A,\bar{A})&=&S_{G/H}(g_{0},A^{\prime},\bar{A}^{\prime})
\nonumber
\\
&=&S_{wznw}(g_{0})-\frac{k}{2\pi}
\int dz^2 Tr[A^{\prime}\epsilon_{+}+\bar{A}^{\prime}\epsilon_{-}
+A^{\prime}g_{0}\bar{A}^{\prime}g_{0}^{-1}],
\label{14}
\end{eqnarray}
since from the graded structure 
$Tr(A\bar{\partial}g_{0}g_{0}^{-1})=Tr(\bar{A}g_0^{-1}\partial g_0)=0$.

After integrating the quadratic terms in the functional integral
\begin{equation}
Z=\int DAD\bar{A}\exp (-F) \quad 
\sim  \quad \exp (-S_{eff}),
\label{15}
\end{equation}
where
\begin{equation}
F=-\frac{k}{2\pi}\int Tr(A^{\prime}\epsilon_{+}+\bar{A}^{\prime}\epsilon_{-}
+A^{\prime}g_{0}\bar{A}^{\prime}g_{0}^{-1}),
\label{16}
\end{equation}
we find the effective action
\begin{eqnarray}
S_{eff}=S_{WZNW}(g_{0})-\frac{k}{2\pi}
\int Tr(\epsilon_{+}g_{0}\epsilon_{-}g_{0}^{-1}).
\label{17}
\end{eqnarray}
\vskip.35cm
\noindent {  1. ABELIAN TODA MODELS.}
 Consider the grading operator
\begin{eqnarray}
Q=\sum_{i=1}^{r}\frac{2\lambda_{i}\cdot H}{\alpha_{i}^{2}},
\label{18}
\end{eqnarray}
where $r=rank \lie$, $H$ is the Cartan subalgebra of $\lie$,
$\alpha_{i}$ is the $i^{th}$ simple root of $\lie$ and $\lambda_{i}$ is the
$i^{th}$ fundamental weight of $\lie$, which satisfies
\begin{eqnarray}
\frac{2\lambda_{i}\cdot \alpha_{j}}{\alpha_{i}^{2}}=\delta_{ij}.
\nonu
\end{eqnarray}
In this case, the Abelian subalgebra of $\lie$, of grade zero, is
\begin{eqnarray}
\lie _{0}=u(1)^{r}=\{ h_{1},h_{2},...,h_{r}\},
\label{20}
\end{eqnarray}
where $h_{i}$, defined as
$ h_{i}=\frac{2\alpha_{i}\cdot H}{\alpha_{i}^{2}} $,
 and satisfy
$[h_{i},E_{\alpha_{j}}]=k_{ji}E_{\alpha_{j}} $;
 $k_{ji}$ denote  the Cartan matrix of $\lie$. Taking all
this into account, the group element in $G_{0}$ may be parameterized as
$ g_{0}=\exp (\phi_{i}h_{i}) $.

The most general constant generators of grade $\pm 1$ are
\begin{eqnarray}
\epsilon_{\pm}=\sum_{i=1}^{r}c_{\pm i}E_{\pm \alpha_{i}}
\label{24}
\end{eqnarray}
and the potential term may be evaluated as
\begin{eqnarray}
Tr(\epsilon_{+}g_{0}\epsilon_{-}g_{0}^{-1})
=\sum_{i=1}^{r}\frac{2}{\alpha_{i}^{2}}c_{i}c_{-i}\exp (k_{ij}\phi_{j}).
\label{25}
\end{eqnarray}
Therefore, the action is given by
\begin{eqnarray}
S_{G/H}=\int d^2z \( \partial \phi_{i}\bar{\partial}\phi_{j}Tr(h_{i}h_{j})
-\sum_{i=1}^{r}\frac{2}{\alpha_{i}^{2}}c_{i}c_{-i}\exp (k_{ij}\phi_{j}) \).
\label{26}
\end{eqnarray}

At this point, let us remark that if $c_{a}=c_{-a}=0$, then
\begin{eqnarray}
\epsilon_{\pm}=\sum_{i\neq a}c_{\pm i}E_{\pm \alpha_{i}}.
\nonu  
\end{eqnarray}
The structure of this Abelian Toda model is such that, by a change of variables
\begin{eqnarray}
g_{0}=\exp (\sum_{i=1}^{a-1}\varphi_{i}h_{i}+\varphi \lambda_{a}\cdot H
+\sum_{i=a+1}^{r}\varphi_{i}h_{i}),
\label{28}
\end{eqnarray}
the action decomposes into two Toda models, together with a free field,
associated to the generator $g_{0}^{0}=\lambda_{a}\cdot H$, such that
$[g_{0}^{0},\epsilon_{\pm}]=0$. Hence, the action can be written as
\begin{eqnarray}
S_{G/H}=S^{(1)}_{G/H}(\varphi_{1},\varphi_{2},...,\varphi_{a-1})
+S^{(2)}_{G/H}(\varphi_{a+1},\varphi_{a+2},...,\varphi_{r})
+{1\o 2}\int d^2z\partial \varphi \bar{\partial}\varphi ,
\nonu 
\end{eqnarray}
where $S^{(1)}$ and $S^{(2)}$ denote the abelian Toda models associated
 to the
decomposed Dynkin diagram of $G$, by deleting the $a^{th}$ point.
\vskip.5cm
 \noindent {  2. NON-ABELIAN TODA MODELS.}
  Consider now the grading operator
\begin{eqnarray}
Q_{a}=\sum_{i\neq a}\frac{2\lambda_{i}\cdot H}{\alpha_{i}^{2}},
\label{30}
\end{eqnarray}
which determines the non-Abelian subalgebra of grade zero
\begin{eqnarray}
\lie_{0}=u(1)^{r-1}\otimes sl(2).
\label{31}
\end{eqnarray}
The group element of grade zero may be then, parameterized as
\begin{eqnarray}
g_{0}=\exp (\chi E_{-\alpha_{a}})
\exp (\sum_{i=1}^{r}\phi_{i}h_{i})\exp (\psi E_{\alpha_{a}}).
\label{32}
\end{eqnarray}
Now, take
\be
\epsilon_{\pm}=\sum_{i\neq a}c_{\pm i}E_{\pm \alpha_{i}}, \quad \quad [\epsilon
_{\pm} , \lambda_{a}\cdot H ] = 0
\label{33}
\ee
such that
\begin{eqnarray}
Tr(\epsilon_{+}g_{0}\epsilon_{-}g_{0}^{-1})=\sum_{i\neq a}
\frac{2}{\alpha_{i}^{2}}c_{i}c_{-i}\exp (\sum_{j\neq a}k_{ij}\phi_{j}),
\nonu 
\end{eqnarray}
 as before. The pure $WZNW$-action for $g_{0}$ acquires a contribution from
the non-abelian structure, 
\begin{eqnarray}
S_{WZNW}=\int d^2z \( \sum_{i,j=1}^{r}\partial \phi_{i}\bar{\partial}\phi_{j}
Tr(h_{i}h_{j})+2\partial \chi \bar{\partial}\psi \exp (k_{ai}\phi_{i}) \).
\label{35}
\end{eqnarray}

Notice that an attempt to decouple a free field from the action by changing the
variables to
\begin{eqnarray}
g_{0}=\exp (\chi E_{-\alpha})
\exp (\sum_{i=1}^{a-1}\varphi_{i}h_{i}+\varphi \lambda_{a}\cdot H
+\sum_{i=a+1}^{r}\varphi_{i}h_{i})\exp (\psi E_{\alpha_{a}} )
\label{36}
\end{eqnarray}
decouples only part of the kinetic term, leading to
\begin{eqnarray}
{\cal L}_{WZNW}(g_{0})=\sum_{i,j\neq a}\partial \phi_{i}\bar{\partial}\phi_{j}
Tr(h_{i}h_{j})+\partial \varphi \bar{\partial}\varphi
+2\partial \chi \bar{\partial}\psi
\exp (\sum_{i\neq a}k_{ai}\varphi_{i}+\varphi).
\nonu 
\end{eqnarray}

It will become clear that the elimination of field $\varphi$ is responsible for
the non-trivial singular metric (of black hole-type) in the string action. 
In order to
gauge away the extra degree of freedom, we need to modify the action to become 
 invariant under transformations within the subspace $ \lie _0^0 = \lambda_a
 \cdot H $ of grade zero (axial gauging), i. e.,
\begin{eqnarray}
g_{0}\longrightarrow g_{0}^{\prime}=\alpha_{0}g_{0}\alpha_{0},
\label{38}
\end{eqnarray}
where $\alpha_{0}=\exp (\varphi \lambda_{a}\cdot H)$. Following the same line of
reasoning as before, we now introduce the new gauge fields
$A=A_{-}+A_{0}$ ,
$ \bar{A}=\bar{A}_{+}+\bar{A}_{0} $,
where $A_{0},\bar{A}_{0}\in \lie_{0}^{0}$ transform as
\begin{eqnarray}
A\longrightarrow A^{\prime}=\alpha_{0}^{-1}A\alpha_{0}
+\alpha_{0}^{-1}\partial \alpha_{0}, \quad \quad 
\bar{A}\longrightarrow \bar{A}^{\prime}=\alpha_{0}\bar{A}\alpha_{0}^{-1}
+\bar{\partial}\alpha_{0}\alpha_{0}^{-1},
\nonu 
\end{eqnarray}
\begin{eqnarray}
A_{0}\longrightarrow A_{0}^{\prime}=A_{0}+\alpha_{0}^{-1}\partial \alpha_{0},
\quad \quad 
\bar{A}_{0}\longrightarrow \bar{A}_{0}^{\prime}=\bar{A}_{0}
+\bar{\partial}\alpha_{0}\alpha_{0}^{-1}.
\nonu 
\end{eqnarray}

The invariant action, under transformations generated by
$H_{(-,0)}$ (in left) and $ H_{(+,0)}$ (in right) is given by
\begin{eqnarray}
S_{G/H}(g,A,\bar{A})&=&S_{WZNW}(g)-\frac{k}{2\pi}\int d^2z 
Tr \( A(\bar{\partial}gg^{-1}-\epsilon_{+}) \right.
\nonumber
\\
&+& \left. \bar{A}(g^{-1}\partial g-\epsilon_{-})+
Ag\bar{A}g^{-1}+A_{0}\bar{A}_{0} \) .
\label{45}
\end{eqnarray}

The redundant fields, corresponding to the nilpotent subalgebras
$\lie_{<0},\;\; \lie_{>0}$ and to $\lie_{0}^{0}$, may be eliminated by
subsequent gauge transformations, generated by $\alpha_{-}$, $\alpha_{+}$ and
$\alpha_{0}$, respectively. Therefore, we arrive at
\begin{eqnarray}
S_{G/H}(g,A,\bar{A})=S_{G/H}(g_{0}^{f},A^{\prime},\bar{A}^{\prime}),
\label{46}
\end{eqnarray}
where
\begin{eqnarray}
g_{0}^{f}=\exp (\chi E_{-\alpha_{a}})\exp (\sum_{i\neq a}\phi_{i}h_{i})
\exp (\psi E_{\alpha_{a}}).
\label{47}
\end{eqnarray}
Due to  the  trace properties 
\br
&&{\rm Tr}A \bar{\partial}g_{0}^f{(g_{0}^f)}^{-1}=
{\rm Tr}A_{0}\bar{\partial}g_{0}^f{(g_{0}^f)}^{-1},
\quad \quad
{\rm Tr}\bar{A}{(g_{0}^f)}^{-1}\partial g_{0}^f=
{\rm Tr}\bar{A}_{0}{(g_{0}^f)}^{-1}\partial g_{0}^f
\nonumber
\\
&&{\rm Tr}Ag_{0}^f\bar{A}^{\prime}{(g_{0}^f)}^{-1}=
{\rm Tr}A_{0}g_{0}^f\bar{A}_0{(g_{0}^f)}^{-1}+
{\rm Tr}A_{-}g_{0}^f\bar{A}_{+}{(g_{0}^f)}^{-1}.
\nonumber
\er
the action decomposes into three parts, i. e.,
\begin{eqnarray}
S_{G/H}=S_{WZNW}(g_{0}^{f})+F_{0}+F_{\pm},
\label{48}
\end{eqnarray}
where
\begin{equation}
F_{0}=-\frac{k}{2\pi}\int d^2z Tr\( A_{0}\bar{\partial}g_{0}^{f}({g_{0}^{f}})^{-1}
+\bar{A}_{0}({g_{0}^{f}})^{-1}\partial g_{0}^{f}
+A_{0}g_{0}^{f}\bar{A}_{0}({g_{0}^{f}})^{-1}+A_{0}\bar{A}_{0} \),
\nonu 
\end{equation}
\begin{eqnarray}
F_{\pm}=-\frac{k}{2\pi}\int dz^2 Tr\( A_{-}\epsilon_{+}+\bar{A}_{+}\epsilon_{-}
+A_{-}g_{0}^{f}\bar{A}_{+}({g_{0}^{f}})^{-1} \),
\nonu 
\end{eqnarray}
and the functional integral now factorizes into 
\begin{eqnarray}
Z=\int DA_{0}D\bar{A}_{0}\exp (-F_{0})\int DA_{-}D\bar{A}_{+}\exp (-F_{\pm}).
\label{51}
\end{eqnarray}

Explicitly making use of the given parameterization for $g_{0}^{f}$ and
writing the gauge fields as $A_{0}=a_{0}(z,\bar z)\lambda_{a}\cdot H$,
$\bar{A}_{0}=\bar{a}_{0}(z,\bar z)\lambda_{a}\cdot H$, we find
\begin{eqnarray}
F_{0}=-\frac{k}{2\pi}\int dz^2 [\frac{a_{0}\bar{a}_{0}}{2\lambda_{a}^{2}}\Delta_{a}
-(a_{0}\chi \bar{\partial}\psi +\bar{a}_{0}\psi \partial \chi )
\exp (k_{ai}\phi_{i})],
\label{52}
\end{eqnarray}
where $
\Delta_{a}=1+\frac{\chi \psi \exp (k_{ai}\phi_{i})}{2\lambda_{a}^{2}} $.

The total effective action, then, becomes
\begin{eqnarray}
S_{eff}&=&-\frac{k}{2\pi}\int dz^2 \( \frac{1}{2}\partial 
\phi_{i}\bar{\partial}\phi_{j}
Tr(h_{i}h_{j})
+\frac{\partial \chi \bar{\partial}\psi}{\Delta_{a}}
\exp (k_{ai}\phi_{i}) \right. \nonu \\
 &-&\left. \sum_{i,j\neq a}^{r}\frac{2}{\alpha_{i}^{2}}
c_{i} c_{-i}\exp (-k_{ij}\phi_{j}) \) .
\label{54}
\end{eqnarray}

The non-symmetric term may be written as
\begin{eqnarray}
\partial \chi \bar{\partial}\psi
=g^{\mu \nu}\partial_{\mu}\chi \partial_{\nu}\psi
-\epsilon_{\mu \nu}\partial_{\mu}\chi \partial_{\nu}\psi .
\nonu
\end{eqnarray}
 and we can rewrite the effective action ( discarding the total derivative
 term) in the form 
\begin{eqnarray}
&&S_{eff}=\int d^2z \( \frac{1}{2}g^{\mu \nu} \partial _{\mu}
 \phi_{i}{\partial _{\nu}}\phi_{j}Tr(h_{i}h_{j})
+g^{\mu \nu}\partial_{\mu}\chi \partial_{\nu}\psi \frac{\exp (k_{ai}\phi_{i})}
{\Delta_{a}} \right.
\nonumber
\\
&-&\left. \frac{1}{2}\epsilon_{\mu \nu}k_{ai}\partial_{\mu}\phi_{i}
(\chi \partial_{\nu}\psi -\psi \partial_{\nu}\chi )\frac{\exp (k_{aj}\phi_{j})}
{\Delta_{a}}-\sum_{i,j\neq a}^{r}\frac{2}{\alpha_{i}^{2}}c_{i}c_{-i}\exp
(-k_{ij}\phi_{j}) \) .
\label{57}
\end{eqnarray}

 As we shall demonstrate below (see \cite{GSZ1}, \cite{GSZ2}) 
 the symmetries of the non-Abelian Toda model (\ref{57}), are generated by
non-local algebraic structures due to the gauge invariance within the zero
grade subspace.

Let us remark that for $\lie =sl(2)$ and for $\phi_{i}=0$ the action
coincides with Witten's black hole action \cite{Witten}, i. e.,
\begin{eqnarray}
S=\int dz^2 g^{\mu \nu}\frac{\partial_{\mu}\chi \partial_{\nu}\psi}{1+\chi \psi}.
\nonu 
\end{eqnarray}
\vskip.35cm
\noindent { 3. NO-TORSION THEOREM.}
  We now discuss the {\it generalized} non-Abelian Toda model, as well as the
``no-torsion theorem''. To do so, first we take the very same grading operator
\begin{eqnarray}
Q_{a}=\sum_{i\neq a}^{r}\frac{2\lambda_{i}\cdot H}{\alpha_{i}^{2}}
\nonu 
\end{eqnarray}
and consider {\it the most general constant generators} of grade $\pm 1$, i. e.,
\begin{eqnarray}
\epsilon_{\pm}=\sum_{i\neq a}^{r}c_{\pm i}E_{\pm \alpha_{i}}
+b_{\pm}E_{\pm (\alpha_{a}+\alpha_{a+1})}
+d_{\pm}E_{\pm (\alpha_{a}+\alpha_{a-1})}.
\label{60}
\end{eqnarray}

It is clear that if $c_{\pm i},b_{\pm},d_{\pm}\neq 0$, there shall be {\it no}
$g_{0}^{0}$ {\it commuting} with $\epsilon_{\pm}$, since  that 
require an orthogonal direction to all roots appearing in $\epsilon_{\pm}$. 
These are the generalized {\it non-singular } NA-Toda models of ref.
\cite{ora}.  
The NA-Toda  models of singular metric $G_{ij}(X)$   
correspond to the cases when $g_0^0 = U(1)$ and we {\it impose}
 it as a subsidiary
constraint \footnote{If we leave $g_{0}^{0}$ unconstrained the resulting
 model belongs again to the  non singular NA-Toda class of models \cite{ora}.}.  Depending
 upon the choice of the constants $c_{\pm i}, b_{\pm}$ and  $d_{\pm}$ we
 distinguish four families of {\it singular} NA-Toda models:
\vskip.3cm
(i)$b_{\pm}=d_{\pm}=0$, $g_{0}^{0}=\frac{2\lambda_{a}\cdot H}{\alpha_{a}^{2}}$;

(ii)$c_{\pm (a-1)}=c_{\pm (a+1)}=0$,
$g_{0}^{0}=\frac{2\lambda_{a}\cdot H}{\alpha_{a}^{2}}
-\frac{2\lambda_{a-1}\cdot H}{\alpha_{a-1}^{2}}
-\frac{2\lambda_{a+1}\cdot H}{\alpha_{a+1}^{2}}$;

(iii)$c_{\pm (a+1)}=d_{\pm}=0$,
$g_{0}^{0}=\frac{2\lambda_{a}\cdot H}{\alpha_{a}^{2}}
-\frac{2\lambda_{a+1}\cdot H}{\alpha_{a+1}^{2}}$;

(iv)$b_{\pm}=c_{\pm (a-1)}=0$,
$g_{0}^{0}=\frac{2\lambda_{a}\cdot H}{\alpha_{a}^{2}}
-\frac{2\lambda_{a-1}\cdot H}{\alpha_{a-1}^{2}}$.
\vskip.3cm
Of course, if $c_{\pm j} = 0$, $j\neq a, a\pm 1$, we find $g_0^0 = \lambda_j
\cdot H$.  However, since $[\lambda_j \cdot H, E_{\pm \alpha_a}] = 0 $,  there
will be no singular metric  present and this case shall be negleted.  
Cases (i) and (ii) are equivalent, since they are related by the Weyl
 reflection
$ \sigma_{\alpha_{a}}(\alpha_{a\pm 1})=\alpha_{a}+\alpha_{a\pm 1}$ and the
corresponding 
 fields are related by non-linear change of the variables. This case has
already been discussed in refs. \cite{GSZ1} and \cite{GSZ2}, and  shown
 to present always the {\it antisymmetric term},
originated by the presence of $e^{k_{ai}\varphi_i}$ in $\Delta_a$  and in the
kinetic trem as well.
Since we are removing all dependence in $g_{0}^{0}$, when parameterizing
$g_{0}^{f}$, cases (iii) and (iv) may be studied together with
\begin{eqnarray}
g_{0}^{f}=\exp (\chi E_{-\alpha_{a}})
 \exp (   \Phi (H))\exp (\psi E_{\alpha_{a}})
 \label{63}
 \end{eqnarray}
where $ \Phi (H) =\sum_{i=1}^{a-2}\varphi_{i}h_i
+ \chi_- \varphi_{-}H + \chi_{+}\varphi_{+}H
+\sum_{i=a+2}^{r}\varphi_{i}h_i $,

\begin{eqnarray}
\chi_{-}^{iii}=\alpha_{a-1}+\alpha_{a}, \quad \quad \chi_{+}^{iii}=\alpha_{a+1},
\label{64}
\end{eqnarray}
\begin{eqnarray}
\chi_{-}^{iv}=\alpha_{a-1},\quad \quad \chi_{+}^{iv}=\alpha_{a}+\alpha_{a+1}
\label{66}
\end{eqnarray}
for cases $iii$ and $iv$ respectively, and 
$g_{0}^{0}=y\cdot H$,
such that
$ Tr(\chi_{\pm}Hg_{0}^{0})=0 $.
It is straightforward to evaluate the integral
\begin{eqnarray}
F_{0}=-\frac{k}{2\pi}\int [2y^{2}a_{0}\bar{a}_{0}\Delta_{a}
-\frac{2}{\alpha_{a}^{2}}(a_{0}\chi \bar{\partial}\psi
+\bar{a}_{0}\psi \partial \chi )\exp (\Phi (\alpha_{a}))],
\nonu 
\end{eqnarray}
where
\begin{eqnarray}
\Phi (\alpha_{a})=\sum_{i=1}^{a-2}k_{ai}\varphi_{i}
+\alpha_{a}\cdot \chi_{-}\varphi_{-}+\alpha_{a}\cdot \chi_{+}\varphi_{+}
+\sum_{i=a+2}^{r}k_{ai}\varphi_{i}
\nonu 
\end{eqnarray}
and
$ \Delta_{a}=1+\frac{\chi \psi \exp (\Phi (\alpha_{a}))}{2y^{2}} $.
The corresponding effective action takes the form 
\begin{eqnarray}
S &=& \int d^2z\( g^{\mu \nu} \partial _{\mu}\varphi_i \partial _{\nu} \varphi_jTr(
h_ih_j) + {{exp (\Phi(\alpha_a ))}\over {y^2 \Delta_a}} g^{\mu \nu }\partial
_{\mu} \psi \partial _{\nu} \chi \right.  \nonumber \\
&-& \left. {1 \over 2} \epsilon _{\mu \nu } \partial _{\mu} \Phi (\alpha _a) (\psi
\partial _{\nu} \chi - \chi \partial _{\nu} \psi ){{exp (\Phi (\alpha_a))}\over
{\Delta _a}} - V \)
\label{72a}
\end{eqnarray}
where the potential is given in terms of the constant elements $\epsilon _{\pm
} $ and $\lie_0^f$ as  
$ V =  \; Tr [\epsilon_+ g_0^{f} \epsilon_- {g_0^{f}}^{-1} ]$.
Now, if we consider Lie algebras whose Dynkin diagrams connect only nearest
neighbours, i. e.,
\begin{eqnarray}
\Phi (\alpha_{a})=\alpha_{a}\cdot \chi_{-}\varphi_{-}
+\alpha_{a}\cdot \chi_{+}\varphi_{+},
\label{73}
\end{eqnarray}
then the ``no-torsion condition'' implies $\Phi (\alpha_{a})=0$.

Considering case (iii), we have
\begin{eqnarray}
\alpha_{a}\cdot \chi_{-}=\alpha_{a}\cdot (\alpha_{a-1}+\alpha_{a})=0,
\label{74}
\end{eqnarray}
\begin{eqnarray}
\alpha_{a}\cdot \chi_{+}=\alpha_{a}\cdot (\alpha_{a+1})=0.
\label{75}
\end{eqnarray}
In this case, the only solution for both equations above is to take $a=r$ (in
such a way that $\alpha_{r+1}=0$) and ${G}=B_{r}$ (so that
$\alpha_{r-1}\cdot \alpha_{r}=-\alpha_{r}^{2}=-1$). This is precisely the case
proposed by Leznov and Saveliev \cite{Leznov} and subsequently discussed by 
Gervais and Saveliev \cite{Gervais} and also by Bilal \cite{Bilal}, for the  
particular case of $B_{2}$.

For case (iv), the ``{\it no-torsion condition}'' requires that
\begin{eqnarray}
\alpha_{a-1}\cdot \alpha_{a}=0, \quad \quad 
\alpha_{a}\cdot (\alpha_{a}+\alpha_{a+1})=0,
\nonu 
\end{eqnarray}
which are satisfied by $a=1$ and also by ${G}=C_{2}$, since
$\alpha_{a-1}=0$ and also $\alpha_{1}\cdot \alpha_{2}=-\alpha_{1}^{2}=-1$,
respectively.

In general, the ``{\it no-torsion condition}'', i. e.,
$  \Phi (\alpha_{a})=0  $,
may be expressed in terms of the structure of the co-set
$G_{0}/{G}_{0}^{0}=\frac{u(1)^{r-1}\otimes sl(2)}{u(1)}$. The
crucial ingredient for the appearence of $\Phi (\alpha_{a})$ arises from the
conjugation
\begin{eqnarray}
Tr(A_{0}g_{0}^{f}\bar{A}_{0}({g_{0}^{f}})^{-1}+A_{0}\bar{A}_{0})=2\lambda_{a}^{2}
\left( 1+\frac{2}{\alpha_{a}^{2}}
\frac{\chi \psi \exp (\Phi (\alpha_{a}))}{2\lambda_{a}^{2}}\right).
\nonu 
\end{eqnarray}

Henceforth, if all generators belonging to the Cartan subalgebra 
parameterizing $g_{0}^{f}$ commute with $E_{\pm \alpha_{a}}$, then
$\Phi (\alpha_{a})=0$, and therefore the structure of the co-set
\begin{eqnarray}
\frac{{G}_{0}}{{G}_{0}^{0}}=\frac{u(1)^{r-1}\otimes sl(2)}{u(1)}
=u(1)^{r-1}\otimes \frac{sl(2)}{u(1)}
\label{80}
\end{eqnarray}
is the general condition for the {\it absence} of the {\it antisymmetric}
 term in the
action.
\vskip.35cm
 {4. ALGEBRA OF SYMMETRIES.}
 The symmetries of the action (\ref{72a}) is expressed in terms of
  its algebra of conserved currents. 
Those are obtained from the Hamiltonian reduction of the WZW model when
particular set of constraints and gauge fixing conditions are implement. 
 The algebra of the remaining currents can be
derived using Dirac brackets or following the method employed in refs. 
\cite{GSZ1} and \cite{GSZ2}, where the explicit form of the algebra of 
symmetries $V_{n+1}^{(1,1)}$  for $a=1$ and $\lie = A_n$ (cases $i$ and $ii$ ) were derived . 
 For $B_2$ we consider $Q_1 = {{2 \lambda_1 H}
\over {\alpha_1^2}}$ and the two Non-Abelian Toda models  can be defined by
choosing $\epsilon_{\pm}^{i} = E_{\pm \alpha_1}$ and 
$\epsilon_{\pm}^{iii} = E_{\pm (\alpha_1+ \alpha_2)}$.  The remaining currents
are $T_{i} = J_{\alpha_1}$, $V^+_{i} = {\sqrt 2} J_{\alpha_1 + \alpha_2}$ and 
 $V^-_{i} = -{\sqrt 2} J_{- \alpha_2}$ and 
$T_{iii} = J_{\alpha_1 + \alpha_2}$, $V^+_{iii} = 
{{1\over {\sqrt 2}}} J_{\alpha_1 + 2\alpha_2}$ and 
 $V^{-}_{iii} = -{1\over {\sqrt 2}} J_{ \alpha_1}$ respectively.  Their algebra
 is given by 
\be
\{T(\sigma), T(\sigma^{\prime})\}
 =  2T(\sigma^{\prime}) 
\delta^{\prime}(\sigma -\sigma^{\prime})  - 
T^{\prime}(\sigma^{\prime})\delta (\sigma -\sigma^{\prime})-
{1\over 2}\delta^{\prime \prime \prime }(\sigma -\sigma^{\prime})
\nonu
\ee
\be
\{T(\sigma), V^{\pm}(\sigma^{\prime})\} =  sV^{\pm}(\sigma^{\prime})
\delta^{\prime}(\sigma -\sigma^{\prime})- {V^{\pm}}^{\prime}(\sigma^{\prime})
\delta (\sigma -\sigma^{\prime}) \nonumber 
\ee
\be
\{V^{\pm}(\sigma), V^{\pm}(\sigma^{\prime})\} = 
tV^{\pm}(\sigma) V^{\pm}(\sigma^{\prime}) \epsilon (\sigma -\sigma^{\prime})
\label{algebra}
\ee
 for $s= {3\over 2}$ and $2$, $t= {1\over 4}$ and $1$ corresponding to cases
 $(i)$ and $(iii)$ respectively.  We also have
 
 \br
\{V_{i}^{\pm}(\sigma), V^{\mp}_i(\sigma^{\prime})\} &= & 
-{1\over 4}(V^{\pm}_i(\sigma) V^{\mp}_i(\sigma^{\prime})  -
 2 V^{\pm}_i(\sigma)^{\prime} V^{\mp}_i(\sigma))
\epsilon (\sigma -\sigma^{\prime}) \nonumber \\
&\pm & T(\sigma^{\prime})
\delta (\sigma -\sigma^{\prime}) 
\mp \delta^{\prime \prime} (\sigma -\sigma^{\prime})
\label{VVi}
\er
 and 
\br
\{V^{\pm}_{iii}(\sigma), V^{\mp}_{iii}(\sigma^{\prime})\} &= & 
-(V^{\pm}_{iii}(\sigma) V^{\mp}_{iii}(\sigma^{\prime})  -
 2 V^{\pm}_{iii}(\sigma)^{\prime} V^{\mp}_{iii}(\sigma))
\epsilon (\sigma -\sigma^{\prime}) \nonumber \\
&- & T^{\prime}(\sigma^{\prime})\d (\s - \s^{\pr}) + 2 T(\sigma^{\prime})
\delta^{\prime} (\sigma -\sigma^{\prime}) - 
{1\over 2} \delta^{\prime \prime \pr} (\sigma -\sigma^{\prime})
\label{VViii}
\er
for cases $(i)$ and $(iii)$ respectively.  
 The nonlocal V-algebra of the $B_2$ (case $iii$)
  was derived in
\cite{Bilal} by explicit construction of the
 conserved currents.  The algebra of symmetries
  of the $B_3$ (case
$iii$) {\it torsionless} NA-Toda model is an example 
of a new type of {\it nonlocal quadratic} algebra of
 VW-type which mixes the features of V- and W-algebras.  It is
generated by two nonlocal currents 
$V^{\pm}$ of spin 3 and two local, $T$ and $W$ of spin 2 
and 4 respectively (all $V^{\pm}$ and $W$ are primary fields):   
  \be
  \{V^{\pm}(\s ), V^{\pm}(\s ^{\pr})\} =
  {1\o 2}V^{\pm}(\s ) V^{\pm}(\s ^{\pr})\eps (\s - \s ^{\pr})
 \nonu
\ee
\br
 \{V^{\pm}(\s ), V^{\mp}(\s ^{\pr})\} &=&-{1\o 2}V^{\pm}(\s )
 V^{\mp}(\s ^{\pr})\eps (\s - \s ^{\pr}) \nonumber \\
&+& ({1\o 5} T^{\pr \pr \pr }(\s) 
- {4 \o 25}(T^2(\s))^{\pr}
 - W^{\pr}(\s) ) \d (\s - \s ^{\pr}) \nonumber \\
&+& ({9\o 10} T^{\pr \pr}(\s) - 
{8\o 25}T^2(\s) -2 W(\s) )   \d ^{\pr} (\s - \s ^{\pr})\nonumber \\
 &+& {3\o 2} T^{\pr}(\s)  \d ^{\pr \pr } (\s - \s ^{\pr})
+ T(\s ) \d ^{\pr \pr \pr } (\s - \s ^{\pr}) - 
{1\o 2} \d ^{(v) } (\s - \s ^{\pr})\nonumber
\er

\br
\{ W(\s ), V^{\mp}(\s ^{\pr})\} & = & -{7\o 5} V^{\pm}(\s)\d ^{\pr \pr \pr } (\s - \s ^{\pr}) - 
{7\o 5} (V^{\pm}(\s) )^{\pr}\d ^{\pr \pr  } (\s - \s ^{\pr})\nonumber \\ & - &
\( {3\o 5}(V^{\pm}(\s ))^{\pr \pr } - 
{{26} \o {25}}T(\s) V^{\pm}(\s)\) \d ^{\pr} (\s - \s ^{\pr})\nonumber \\
&-&\( { 1\o 10 } (V^{\pm}(\s))^{\pr \pr \pr } - 
{1\o 2} T^{\pr}(\s ) V^{\pm} (\s) + 
{9\o {25} } T(\s ) (V^{\pm}(\s))^{\pr} \) \d(\s - \s ^{\pr}) \nonumber
\er
\br
\{W(\s ),W(\s ^{\pr} ) \} & = & 
 -{1\o 20}\d ^{(vii) } (\s - \s ^{\pr}) + {7\o 25}T(\s )\d ^{(v )} (\s - \s ^{\pr})
+ {7\o 25} T^{\pr}(\s) \d ^{(iv) } (\s - \s ^{\pr}) \nonumber \\
& + & \sum_{i=4}^{7} A_i(\s)\d ^{(7-i)}  (\s - \s ^{\pr})
\nonumber 
\er
where
\br
A_4 (\s )
&=& {21 \o 25} T^{\pr \pr}(\s ) - {49 \o 125} T^{2}(\s) + {3\o 5}W(\s) 
  \nonu \\ 
A_5 (\s )
&=& {14 \o 25 }T^{\pr \pr \pr }(\s )- {147 \o 250} (T^2(\s ))^{\pr}  +
 {9\o 10 } W^{\pr}(\s )  \nonumber \\
A_6 (\s )&=&  {1\o 5} T^{(iv)}(\s ) - {113 \o 200} (T^2(\s ))^{\pr \pr } + 
{213 \o 500} T(\s ) T^{\pr \pr }(\s ) + {27 \o
50 } (T^{\pr}(\s ))^2 + {72 \o 625} T^3(\s )  \nonumber \\ 
&+& {1\o 2} W^{\pr \pr }(\s ) - {14 \o 25} T(\s ) W(\s ) - 3 V^{+}(\s)
V^{-}(\s) \nonumber \\
A_7 (\s )
&=& {3\o 100} T^{(v)}(\s ) - {9 \o 100} (T^2(\s ))^{\pr \pr \pr} - 
{3\o 10 }(T(\s) T^{\pr \pr }(\s))^{\pr} + {243 \o
500} T^{\pr}(\s) T^{\pr \pr }(\s) \nonumber \\ 
& + & {81 \o 250}T(\s) T^{\pr \pr \pr} (\s)
 + {9\o 100}(T^3(\s))^{\pr} - {243 \o 5000}
T(\s) (T^2(\s))^{\pr} + {1\o 10 }W^{\pr \pr \pr}(\s)\nonumber \\ & -&
 {7 \o 25}(T(\s) W(\s) )^{\pr}
   - {3 \o 2} (V^{+}(\s) V^{-}(\s))^{\pr}
 \nonumber
 \er
 
 The {\it axionic} $A_3$-NA-Toda model corresponding to 
 $Q= (\lambda_1 + \lambda_3)\cdot H$, $ \eps _{\pm} = E_{\a_1} +
 E_{\a_3}$, and $g_0^0 = \lambda_2 \cdot H$  provides an 
  example of a {\it new algebraic structure} of $UV$-type.  The
 $V_4^{(2,1)}$-algebra  is generated by three nonlocal currents 
 $V^{\pm}$ and $V^0$ of spin 2  and a local  one (stress
 tensor ) $T$ of spin 2:
\br
 \{ V^{\pm }(\sigma ), V^{\pm }(\sigma^{\pr} ) \} & =& {1\o 8} 
 \eps ( \sigma - \sigma^{\pr} )V^{\pm }(\sigma ) V^{\pm }(\sigma^{\pr} ),
 \nonu \\
\{ V^{0}(\sigma ), V^{\pm }(\sigma^{\pr} ) \}   &=&  {1\o 8} 
 \eps ( \sigma - \sigma^{\pr} )V^{0 }(\sigma ) V^{\pm }(\sigma^{\pr} )
\nonu 
\er
\br
\{ V^{0}(\sigma ), V^{0}(\sigma^{\pr} ) \} & =& -{1\o 4} 
\eps ( \sigma - \sigma^{\pr} )[V^{+ }(\sigma ) V^{- }(\sigma^{\pr} ) +
V^{-}(\sigma ) V^{+}(\sigma^{\pr} )] \nonumber \\
&+& 2 T(\sigma ^{\pr} )\pa _{\sigma^{\pr}} \d (\sigma - \sigma^{\pr} ) + 
\d (\sigma - \sigma^{\pr} ) \pa _{\sigma^{\pr}}T(\sigma^{\pr} ) - 
4 \pa ^{3}_{\sigma^{\pr}}\d (\sigma - \sigma^{\pr} )
 \nonumber 
 \er
 \br
 \{ V^{- }(\sigma ), V^{+ }(\sigma^{\pr} ) \} & =& -{1\o 8} 
 \eps ( \sigma - \sigma^{\pr} )[V^{0 }(\sigma ) V^{0 }(\sigma^{\pr} ) +
V^{-}(\sigma ) V^{+}(\sigma^{\pr} )] \nonumber \\
& + & T(\sigma ^{\pr} )\pa _{\sigma^{\pr}} \d (\sigma - \sigma^{\pr} )
+ {1\o 2}\d (\sigma - \sigma^{\pr} ) \pa _{\sigma^{\pr}}T(\sigma^{\pr} )
- 2\pa ^{3}_{\sigma^{\pr}}\d (\sigma - \sigma^{\pr} )
\nonu \\
\label{b}
\er
 By another choice of gauge fixing conditions \cite{GSSZ1}
  (by nonlocal gauge transformations  that relates  two
 different gauge fixings) one can transform this algebra
  into nonlocal {\it rational} algebra $VU_4^{(2,1)}$ generated by
the nonlocal $V^{\pm}$ currents and now $V_0$ is local but appears
in the denominator of the r.h.s. of (\ref{b}).  
 The origin of this novelty is that together with the $\eps_{\pm}$ 
 invariant subalgebra $g_0^0$ there exists one
 extra   $\eps_{\pm}$   {\it invariant subalgebra} $K^-$ i.e. 
 \be
 [\eps_{\pm}, E_{-\a_1 -\a_2} - E_{-\a_2 -\a_3}] =0
 \nonu
 \ee
 It is important to note that relaxing the $g_0^0$ constraint, 
 the algebra of symmetries of the corresponding
 {\it nonsingular} NA-Toda is generated by three local currents $V^{\pm}, 
 T$ of spin 2, one nonlocal $V_0$ ($s=2$) and one local spin 1 current $J$:

 \begin{eqnarray}
\{ V^0(\sigma ),J(\sigma^{\prime})\}
=\{ V^{\pm}(\sigma ),V^{\pm}(\sigma^{\prime})\} =0,
\quad  
\{ V^0(\sigma ),V^{\pm}(\sigma^{\prime})\}
=\frac{1}{2}V^{\pm}(\sigma )V^0(\sigma^{\prime})
\epsilon (\sigma -\sigma^{\prime}),
\nonumber
\end{eqnarray}
\begin{eqnarray}
\{ J(\sigma ),V^{\pm}(\sigma^{\prime})\}
=\mp \frac{1}{4}V^{\pm}(\sigma )\delta (\sigma -\sigma^{\prime}),
\quad \quad 
\{ J(\sigma ),J(\sigma^{\prime})\}
=\frac{1}{8}\partial_{\sigma}\delta (\sigma -\sigma^{\prime}),
\nonumber
\end{eqnarray}
\begin{eqnarray}
\{ V^{\pm}(\sigma ),V^{\mp}(\sigma^{\prime})\}
&=&-\frac{1}{2}\partial_{\sigma}^{3}\delta (\sigma -\sigma^{\prime})
+{\cal T}(\s )\partial_{\sigma}\delta (\sigma -\sigma^{\prime})
\nonumber
\\
&+&\frac{1}{2}\partial_{\sigma}{\cal T}(\s )
\delta (\sigma -\sigma^{\prime})
-\frac{1}{4}V^0(\sigma )V^0(\sigma^{\prime})\epsilon (\sigma -\sigma^{\prime}),
\nonumber
\end{eqnarray}
\begin{eqnarray}
\{ V^0(\sigma ),V^0(\sigma^{\prime})\}
&=&-\partial_{\sigma}^{3}\delta (\sigma -\sigma^{\prime})
+2{\cal T}(\s )\partial_{\sigma}
\delta (\sigma -\sigma^{\prime})
\nonumber
\\
&+&\partial_{\sigma}{\cal T}(\s )\delta 
(\sigma -\sigma^{\prime})\nonumber
\\
&-&[V^{+}(\sigma )V^{-}(\sigma^{\prime})+V^{+}(\sigma^{\prime})V^{-}(\sigma )]
\epsilon(\sigma -\sigma^{\prime}).
\label{23}
\end{eqnarray}
where ${\cal T} (\s) = T - 4J^2$.
 Again by suitable change of the gauge fixing \cite{GSSZ1},
  this algebra takes the form of rational $U_4^{(1,2)}$-algebra 
 (the current $V^0$ appears in the denominators in the
  $V^+V^-$-Poisson brackets).

  A complete discussion 
 of the symmetries of all grade
  $l=1$ (i.e. $\eps_{\pm} \in \lie_1$), $ Q = \sum_{i \neq a} \lambda_i \cdot H$
  NA-Toda
 models (including the nonsingular 
 cases $g_0^0 =0$) is given in our forthcoming paper \cite{GSSZ1} 
 ( see also ref. \cite{GSSZ2})
 
 \vskip.5cm
{\bf Acknowledgments}  F.E. Mendon\c ca da Silveira thanks FAPESP for financial
support.  G.M.Sotkov thanks FAPESP, IFT-UNESP and
DCP-CBPF for financial support and hospitality. 
 This work was supported in part by CNPq.

\end{document}